%% file: main0WSDM0PreRec.tex
\documentclass[sigconf]{acmart}

\usepackage[utf8]{inputenc} 
\usepackage{url}            
\usepackage{amsfonts}       
\usepackage{xcolor}         

\usepackage{amsmath}
\usepackage{multirow}
\usepackage{threeparttable}
\usepackage{tabularx}
\usepackage{paralist}

\usepackage{bm}
\usepackage{bbm}

\usepackage{amsthm}
\usepackage{booktabs}       
\usepackage{nicefrac}       
\usepackage{microtype}      

\usepackage{wrapfig}

\usepackage{siunitx}

\usepackage{subfig}
\usepackage{graphicx}
\usepackage{arydshln}
\usepackage{dashrule}
\usepackage{enumitem}
\usepackage{afterpage}
\newcommand\blankpage{%
    \null
    \thispagestyle{empty}%
    \addtocounter{page}{-1}%
    \newpage}
\usepackage{xspace}
\newcommand{\model}{\textsc{PreRec}\xspace}
\newcommand{\transformer}{$\text{PreRec}_n$\xspace}
\newcommand{\gru}{ZESRec\xspace}
\newcommand{\sbert}{SBERT\xspace}

\newcommand{\transformerS}{SASRec\xspace}
\newcommand{\gruS}{GRU4Rec\xspace}
\newcommand{\unisrec}{UniSRec\xspace}

\newcommand{\nop}[1]{}

\usepackage[textsize=tiny]{todonotes}

\input{defs.tex}

\AtBeginDocument{%
  }

\copyrightyear{2024}
\acmYear{2024}
\setcopyright{acmlicensed}\acmConference[WSDM '24]{Proceedings of the 17th ACM International Conference on Web Search and Data Mining}{March 4--8, 2024}{Merida, Mexico}
\acmBooktitle{Proceedings of the 17th ACM International Conference on Web Search and Data Mining (WSDM '24), March 4--8, 2024, Merida, Mexico}
\acmPrice{15.00}
\acmDOI{10.1145/3616855.3635779}
\acmISBN{979-8-4007-0371-3/24/03}

\begin{document}

\title{Pre-trained Recommender Systems: A Causal Debiasing Perspective}

\author{Ziqian Lin}
\authornote{Both authors contributed equally to this research.}
\authornote{Work done during the author’s internship at AWS AI Labs.}
\email{zlin284@wisc.edu}
\affiliation{%
  \institution{University of Wisconsin-Madison}
  \country{USA}
}
\author{Hao Ding}
\authornotemark[1]
\email{haodin@amazon.com}
\affiliation{%
  \institution{AWS AI Labs}
  \country{USA}
}

\author{Nghia Trong Hoang}
\email{trongnghia.hoang@wsu.edu}
\authornote{Work done at AWS AI Labs.}
\affiliation{
  \institution{Washington State University}
  \country{USA}
}

\author{Branislav Kveton}
\email{bkveton@amazon.com}
\affiliation{%
 \institution{AWS AI Labs}
 \country{USA}
}

\author{Anoop Deoras}
\email{adeoras@amazon.com}
\affiliation{%
 \institution{AWS AI Labs}
 \country{USA}
}

\author{Hao Wang}
\email{howngz@amazon.com}
\affiliation{%
 \institution{AWS AI Labs}
 \country{USA}
}

\renewcommand{\shortauthors}{Ziqian Lin et al.}


\begin{abstract}
Recent studies on pre-trained vision/language models have demonstrated the practical benefit of a new, promising solution-building paradigm in AI where models can be pre-trained on broad data describing a generic task space and then adapted successfully to solve a wide range of downstream tasks, even when training data is severely limited (e.g., in zero- or few-shot learning scenarios). Inspired by such progress, we investigate in this paper the possibilities and challenges of adapting such a paradigm to recommender systems. 
In particular, we develop a generic recommender that captures universal interaction patterns by training on generic user-item interaction data extracted from different domains, which can then be fast adapted to improve few-shot learning performance in unseen new domains (with limited data). 

However, unlike vision/language data which share strong conformity in the semantic space, universal patterns underlying recommendation data collected across different domains 
(e.g., different countries) 
are often occluded by both in-domain and cross-domain biases implicitly imposed by the cultural differences in their user and item bases. 
Such heterogeneous biases tend to hinder the effectiveness of pre-trained models. 
To address this challenge, we further introduce and formalize a causal debiasing perspective, which is substantiated via a hierarchical Bayesian deep learning model, named \model. Our empirical studies on real-world data show that the proposed model could significantly improve the recommendation performance in zero- and few-shot learning settings under both cross-market and cross-platform scenarios. 
Our code is publicly available at GitHub for reproducibility\footnote{\href{https://github.com/myhakureimu/PreRec}{https://github.com/myhakureimu/PreRec}}.
\end{abstract}



\ccsdesc[500]{Information systems~Recommender systems}

\keywords{Recommender Systems, Pre-Trained Models, Causality, Probabilistic Methods, Bayesian Inference}

\maketitle

\section{Introduction}

Recommender systems (RecSys)
~\cite{gru4rec,sasrec,lmrecsys,pfdai2i,trendrec}
recommend relevant items from a large inventory based on the personal interests of the user, which mitigates the information overload issue on the Internet. They have been proven to be effective on a wide range of application scenarios including e-commerce and online media. An essential problem for RecSys is data scarcity, which restrains the recommendation performance. Recent years have witnessed the great success of the pre-trained language model (PLM) such as BERT~\cite{bert} and GPT~\cite{gpt3} in natural language processing (NLP), which improves the performance on the target domain by leveraging massive training data from other domains. A natural question that arises is whether it is feasible to build a pre-trained RecSys that is generally applicable across different domains.

Traditional approaches, including collaborative filtering-based methods ~\cite{pmf,cdl} and session-based methods ~\cite{gru4rec,sasrec,hrnn}, typically learn item embeddings indexed by domain-specific item IDs (also known as \textit{item ID embeddings}) through interaction data. Such item ID embeddings are transductive and are not generalizable to unseen items, which causes the in-domain item cold-start issue and also becomes a blocker for transferring knowledge from one domain to another. Of course we can infer the embedding of the unseen item based on its metadata, however different domains are likely to possess different sets of item metadata. For example, the item metadata in domain $m$ is \textit{\{director, actor\}} while in domain $n$ is \textit{\{genre, rating\}}, the knowledge that the model learned based on item metadata in domain $m$ can never be transferred to domain $n$ and vice versa. 

A new line of deep learning recommenders, pioneered by collaborative deep learning (CDL)~\cite{cdl} and its variants~\cite{CRAE,ColVAE}, seamlessly incorporate content information into deep recommenders, thereby opening up the possibility of pretraining-based recommenders and significantly alleviating data scarcity (and cold-start) problems in recommender systems. 
Inspired by CDL, we argue that the key to learning a generalizable model lies in capturing knowledge grounded on a universal feature space. In order to bridge the gap between different domains, ZESRec~\cite{ZESRec} first proposed to use generic item textual description (e.g., product description, movie synopsis, and news content) to produce the item universal embedding, while the user universal embedding is computed via a sequential model (Gated Recurrent Units or GRU, Transformer, etc.) aggregating item universal embeddings for items in the user history. Despite the promising results, ZESRec limits itself to conducting pre-training on a single source domain and inference on a single target domain. As follow-up work, UniSRec~\cite{unisrec} further extends support for multi-domain pre-training by designing cross-domain contrastive learning tasks, and evaluate the pre-trained model on multiple target domains. However, UniSRec fails to consider the bias either within each domain or across domains during pre-training, while both types of bias may lead to drift in user interests, item properties, and user behavioral patterns. Training the model on signals mixed with bias and general knowledge without considerations of debiasing will lead to overfitting and impairs the generalizability of the model, as shown in Sec.~\ref{sec:experiment}.

In this work, we {aim to design a universally generalizable recommender that can be pre-trained on multiple source domains and fine-tuned on different target domains. We} start by identifying two types of bias: 
\begin{enumerate}[leftmargin=17pt]
\item \textbf{In-domain bias} considers the biases that take effect within each domain, and one example is the popularity bias that affects not only the item exposure rate, but also the user behavioral pattern due to users tend to follow the majority and interact with trending items (user conformity effect ~\cite{conformity}).
\item \textbf{Cross-domain bias} considers the item bias introduced by the unique domain properties, for instance, a domain launches a promotion campaign for a set of items which influences both item price and user behavioral patterns. Besides, each domain has its characters, which will form its distinctive user community; this causes the shift in user interests across different domains. 
\end{enumerate}

To address the aforementioned challenges, we propose a novel Bayesian deep learning framework, dubbed Pre-trained Recommender Systems (\model), which is equipped with a causal debiasing mechanism and distills general knowledge from massive multi-domain data. {Our model isolates the effects of both in-domain bias and cross-domain bias by explicitly incorporating corresponding bias terms into the modeling during pre-training, and performs debiasing by (1) enforcing causal intervention and neutralizing the cross-domain bias and (2) inferring and removing the in-domain bias during inference in the target domain.} In this way, our model utilizes the bias terms to explain away the variability originating from in-domain and cross-domain dynamics, and is capable of capturing universal item properties and then inferring universal user interests accordingly. In order to adapt to the new domain, \model is enabled to capture the bias in the target domain during fine-tuning and integrate such bias during inference to improve recommendation performance. To summarize our contributions:
\begin{itemize}[leftmargin=15pt]
\item We first identify the in-domain bias and the cross-domain bias existing in the multi-domain data which may potentially undermine the generalizability of the pre-trained RecSys.
\item We design a novel Bayesian deep learning model, named \model, which distills general knowledge from multi-domain data. It captures the in-domain bias and the cross-domain bias, and adds casual intervention for debiasing.   
\item We extensively conduct experiments on datasets collected from different domains and evaluate \model on multiple target domains in three settings including zero-shot, incremental training, and fine-tuning, under both cross-market and cross-platform scenarios. The empirical results demonstrate the effectiveness of our model.
\end{itemize}

\section{Preliminary} \label{sec:preliminary}
In this paper, we focus on training a generalizable recommender system on multiple source domains; to achieve this, the model needs item universal representations and user universal representations. As preliminaries, here we introduce the item universal embedding network (item UEN) and the user universal embedding network (user UEN).


\textbf{Item Universal Embedding Network.}
An item universal embedding network, denoted as $f_\textit{e}(\cdot)$,  generates item $j$'s embedding based on its textual description $\X_j$. The network consists of a pre-trained language model (PLM) to extract generic semantic features from $\X_j$; in our work we adopt a multi-lingual version of Sentence-BERT \footnote{https://huggingface.co/sentence-transformers/stsb-xlm-r-multilingual}~\cite{sbert}, denoted as $f_\textit{BERT}(\cdot)$, followed by a single-layer neural network $f_\textit{NN}(\cdot)$. The tokenizer of Sentence-BERT ingests $\X_j$ to yield a set of tokens $\{t_1, t_2, ..., t_T\}$ before sending the input to the model, formally:
\begin{align}
\m_j = f_\textit{e}(\X_j) = f_\textit{NN}(f_{\textit{BERT}}(\{t_1, t_2, ..., t_T\})), \label{eq:item_uen}
\end{align}
where $\m_j$ represents item universal embedding of item $j$. 

\textbf{User Universal Embedding Network.} 
We denote the item universal embeddings (in chronological order) of historical items for user $i$ as an embedding matrix $\H_i \in \mathbb{R}^{N_u \times B}$, where $N_u$ and $B$ represent the number of items in the history of user $i$ and the hidden dimension of item universal embedding. Note that $\H_i$ is time-sensitive, and we ignore the time indexing for simplicity. A user universal embedding network is essentially a sequential model serving as an aggregation function over $\H_i$:
\begin{align}
\u_i = f_{\textit{seq}}(\H_i), \label{eq:user_uen}
\end{align}
where $\u_i$ stands for user universal embedding of user $i$, $f_{\textit{seq}}(\cdot)$ denotes any type of sequential model, and in this work, we consider a transformer-based recommender ~\cite{sasrec} due to its superior performance.

An existing issue for user UEN is that it cannot generate user embedding for users without interactions. To address this issue, one can pad a learnable dummy item at the front of each user sequence, which can be seen as a prior of user behavioral patterns. In our pre-training stage, we consider user communities in multiple domains, where each domain needs its own prior. We therefore adjust the user UEN to accommodate domain prior: 
\begin{align}
\n_i = f_{\textit{seq}}(\D_k, \H_i), \label{eq:user_emb_net}
\end{align}
where $\D_k \in \mathbb{R}^{B}$ denote the prior of domain $k$, $\n_i$ denotes and the user embedding of user $i$. We assume $\D_k$ is drawn from a zero-mean isotropic multivariate Gaussian distribution. 




\section{Pre-trained Recommender Systems} \label{sec:pretrained_recsys}

In this section, we introduce \model, a flexible hierarchical Bayesian deep learning framework~\cite{BDL,BDLSurvey,CRAE} and can be easily extended to any sequential model (in this work we adopt Transformer). It can be pre-trained on multiple source domains to distill knowledge grounded on a universal feature space; in this work we focus on the textual feature space as a possible instantiation of the universal feature space, but note that our model is generally applicable to any other modalities such as images. In general, \model works in three stages:

\begin{enumerate}[leftmargin=17pt]
\item \textbf{Multi-domain Pre-training:} \model is pre-trained on the data collected from multiple source domains, taking into account both cross-domain and in-domain bias.
\item \textbf{Zero-shot Recommendation:} \model recommends in the target domain without fine-tuning on any target domain interactions. 
\item \textbf{Fine-tuning:} \model is further fine-tuned on the retrieved target domain data, capturing domain bias and popularity bias in the target domain, and adjusting target domain inference accordingly.
\end{enumerate}

\subsection{Model Overview}
\figref{fig:pgm} shows the graphical model for \model. 
(See the supplement for the model architecture from a neural network perspective.) Below we explain its rationale in detail:
\begin{itemize}[leftmargin=15pt]
    \item Variable $\D_k\in\mathbb{R}^B$ represents the (latent) properties of domain $k$ such as the user community, promotion campaigns, as well as website design. 
    \item Variables $\U_i\in\mathbb{R}^B$ and $\H_i\in\mathbb{R}^{N_u \times B}$ represent the genuine interests of user~$i$ and user $i$'s interaction history, respectively. Note that we ignore the time indexing of $\H_i$ for simplicity. 
    \item Variable $\F_j\in\mathbb{R}^C$ represents the popularity factors of item $j$, including three prominent factors: (1) number of interactions of item $j$, (2) number of interactions of all items (traffic volume), and (3) time. We ignore the time indexing of $\F_j$ for simplicity. 
    \item Variables $\X_j$, $\Z_j\in\mathbb{R}^B$, and $\V_j\in\mathbb{R}^B$ represent item $j$'s textual description, item $j$'s popularity properties derived from $\F_j$, and item $j$'s overall properties, respectively. Since a user only interacts with displayed items, we assume item $j$ is exposed. 
    \item Variable $\R_{ijk}\in\{0,1\}$ is the interaction label denoting whether user $i$ interacted with item $j$ in domain $k$. 
    \item Edges $\D_k \to \{\U_i, \V_j, \R_{ijk}\}$: Domain properties may influence user interests, item properties, and user behavioral patterns. 
    \item Edge $\F_j \to \Z_j$: Item popularity factors decide the popularity properties.
    \item Edge $\H_i \to \U_i$: What the user has interacted in the past may affect user's next move (e.g., a user who purchased a cell phone may want to purchase its accessories next). 
    \item Edges $\{\X_j, \Z_j\} \to \V_j$: Item textual descriptions affect item properties and the popularity factors affect item exposure rate. 
    \item Edge $\{\U_i, \V_j, \D_k, \Z_j\}\to\R_{ijk}$: Interaction depends on user interests $\U_i$, item properties $\V_j$, domain properties $\D_k$, and the user conformity effect caused by $\Z_j$.
\end{itemize}
Here $\D_k, \U_i, \V_j$, and $\Z_j$ have the same hidden dimension, and we name them \textit{latent domain embedding}, \textit{latent user embedding}, \textit{latent item embedding}, and \textit{latent popularity embedding}, respectively. We also name $\X_j$ as \textit{item textual description}, and $\Z_j$ as \textit{popularity factors}. {The corresponding conditional probabilities in the PGM are listed in ~\eqnref{eq:map_latent_variables}.} In general, this is a hierarchical Bayesian deep learning (BDL) model~\cite{BDL,BDLSurvey} with the Transformer as the deep component and the probabilistic graphical model in~\figref{fig:pgm} as the graphical component to capture the causal relation.

\begin{figure}[t]
  \begin{center}
    \includegraphics[scale=0.45]{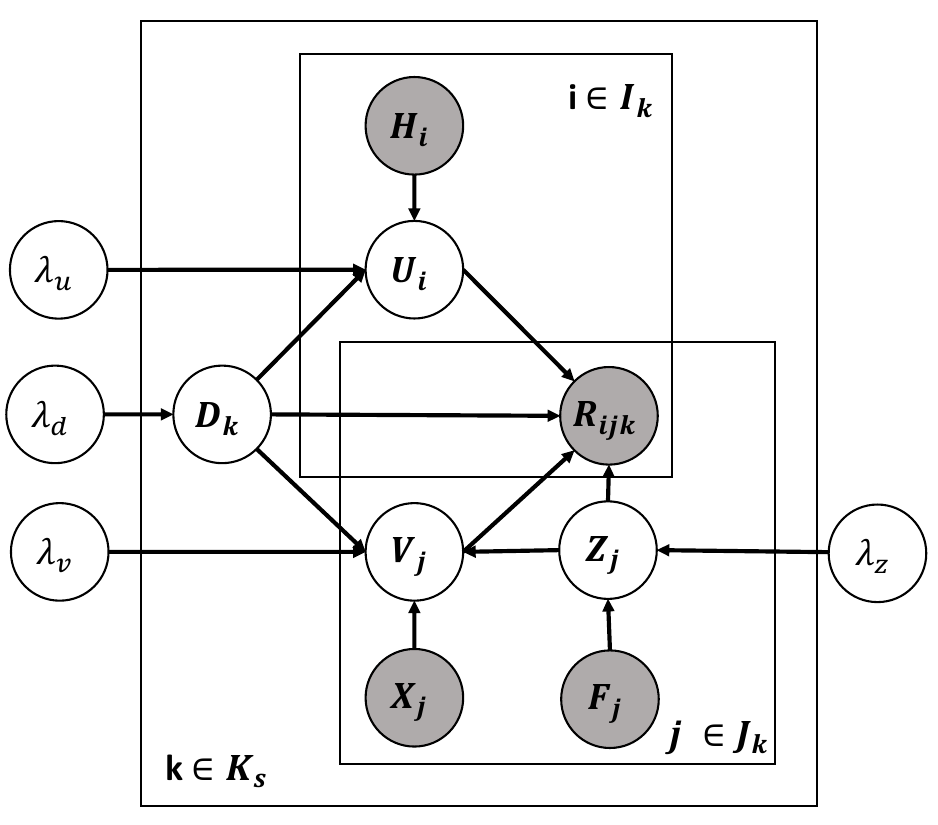}
  \end{center}
  \vskip -0.3cm
  \caption{The probabilistic graphical model (PGM) for our \model. $\U_i$ and $\H_i$ represent user $i$ and corresponding user history; $\V_j$, $\X_j$, and $\Z_j$ represent item $j$, its textual description (e.g., movie synopsis), and its popularity effect; $\F_j$ represents all the prominent factors impact item popularity; $\D_k$ represents domain $k$. $K_s$ represents all the source domains, while $J_k$, $I_k$ represent all the items and all the users in domain $k$, respectively.  $\lambda_u$, $\lambda_v$, $\lambda_d$, $\lambda_z$ are hyperparameters related to distribution variance. }\label{fig:pgm}
  \vskip -0.5cm
\end{figure}

\textbf{Generative Process.} Inspired by CDL~\cite{cdl} and ZESRec~\cite{ZESRec}, the generative process in \figref{fig:pgm} for each domain $k$ in source domains $\mathcal{D}$ is defined as follows:
\begin{enumerate}[leftmargin=15pt]
    \item Draw a latent domain embedding $\D_k \sim \mathcal{N}\left(\0, \lambda_{d}^{-1} \I_{B}\right)$.
    \item For each item $j$:
        \begin{itemize}[leftmargin=4pt]
            \item Draw a latent item offset vector $\boldsymbol\epsilon^v_j \sim \mathcal{N}\left(\0, \lambda_{v}^{-1} \I_{B}\right)$.
            \item Obtain the item universal embedding: $\m_j = f_{e}(\X_j)$.
            \item Draw a latent popularity offset  $\boldsymbol\epsilon^z_j \sim \mathcal{N}\left(\0, \lambda_{z}^{-1} \I_{B}\right)$.
            \item Compute the latent popularity embedding based on popularity factors $\F_j$ directly derived from the data and the latent popularity offset vector: $\Z_j = f_\textit{pop}(\F_j) + \boldsymbol\epsilon^z_j$.
            \item Compute the latent item embedding:
                $
                    \V_j = \boldsymbol\epsilon_j^v +  f_\textit{item}(\D_k, \Z_j, \m_j),
                $ where $f_\textit{item}(\cdot)$ represents the item encoder: $f_\textit{item}(\D_k, \Z_j, \m_j) = \D_k + \Z_j + \m_{j}$.
        \end{itemize}
    
     \item For each user $i$:
     \begin{itemize}[leftmargin=4pt]
         \item Draw a latent user offset vector $\boldsymbol\epsilon^u_i \sim \mathcal{N}\left(\0, \lambda_{u}^{-1} \I_{B}\right)$.   
        \item Obtain the user embedding: $\n_{i} = f_\textit{seq}(\D_k, \H_i)$.
        \item Compute the latent user embedding: $\U_i = \boldsymbol\epsilon^u_i + \n_{i}$.
        \item Compute the recommendation score $\S_{ijk}$ for each tuple $(i, j, k)$,             
        $
            \S_{ijk} = f_\textit{softmax}(\U_{i}^\top \V_j + \D_k \W_d + \Z_j \W_z)
        $, where $\W_d$ and $\W_z$ are trainable parameters, and for user $i$: $\R_{i*k} \sim Cat([\S_{ijk}]_{j=1}^{J_k})$, 
        where $\R_{i*k}\in \{0,1\}^{J_K}$ is a one-hot vector denoting an item ID ($\R_{ijk}=1$ if $j$ is the sampled item index from $Cat([\S_{ijk}]_{j=1}^{J_k})$, and $\R_{ijk}=0$ otherwise); 
        $J_k$ denotes number of items in domain $k$, `$*$' represents the collection of all elements in a specific dimension, $f_\textit{softmax}(\cdot)$ represents the softmax function, and $Cat(\cdot)$ denotes a categorical distribution.
    \end{itemize}
\end{enumerate}

Below we make a few remarks on this generative process. 
\begin{itemize}[leftmargin=15pt]
\item $\D_k$, $\Z_j$, and $\m_j$ in $f_\textit{item}(\D_k, \Z_j, \m_j)$ capture cross-domain bias, in-domain (popularity) bias, and item content (e.g., text description), respectively. 
\item The latent offset vectors $\boldsymbol\epsilon^v_j$, $\boldsymbol\epsilon^z_j$, $\boldsymbol\epsilon^u_i$ accommodate variances in item properties, popularity properties, and user interests. For example, $\ep^v_j$ provides flexibility that the final item vector $\v_j$ can deviate from $f_\textit{item}(\D_k, \Z_j, \m_j)$ and that $\ep^v_j$ can be different for different item $j$. 
\item For zero-shot recommendation (see Sec.~\ref{subsec:zero_shot}), we remove all the latent offset vectors and set $\D_k=\mathbf{0}$ to eliminate noises.
\end{itemize}

\textbf{Computing Popularity Factors $\F_j$.} We now provide the details of computing the popularity factors $\F_j$ and the popularity properties $\Z_j$ for item $j$. Note that $\F_j$ is time-sensitive, and we ignore the time indexing for simplicity. We divide the interaction data $\R$ into a series of equal size time intervals $\mathcal{T} = \{T_1, T_2, ..., T_{\mathcal{T}}\}$. Assuming time $t$, item $j$, and domain $k$ are given, where $t$ falls into the time interval $T_{l+1}$, we first obtain the number of interactions for item $j$ in the former time interval $T_{l}$ (prevent temporal leakage) and denote it as $c_j^{T_l}$, and use $J_k$ to represent items in domain $k$. The popularity factors $\F_j$ is computed as:
\begin{gather*}
   \F_j = [{c_j^{T_l}}/{s_1^{T_l}}, {c_j^{T_l}}/{s_2^{T_l}}, ... ,{c_j^{T_l}}/{s_w^{T_l}}],
\end{gather*} 
where $s_w^{T_l}$ is a normalization term calculated as:
\begingroup\makeatletter\def\f@size{8}\check@mathfonts
\begin{gather*}
   s_w^{T_l} = (\sum\nolimits_{j\in J_k} (c_j^{T_l})^w / |J_k|)^{\frac{1}{w}}
\end{gather*}
\endgroup

We then use the trainable single layer neural network $f_{pop}(\cdot)$ to obtain $\Z_j = f_{pop}(\F_j)$. Note that the $f_{pop}(\cdot)$ is \emph{generally applicable across different domains}. See See the supplement for the design philosophy and visualization of computing popularity properties.



\subsection{Multi-domain Pre-training} \label{subsec:pre-training}
For multi-domain pre-training, our unique challenge is to leverage training data from different source domains \emph{without introducing bias}. The sources of bias can be broadly summarized into two categories: (1) in-domain bias caused by variances within each domain such as popularity bias~\cite{hrnn, zhang2021causal, zheng2021disentangling}, and (2) cross-domain bias originated from dynamics across different domains, including changes in user community, item catalogue, promotion campaigns, etc.


\textbf{In-domain Bias and Explicit Confounders.} 
For in-domain bias, we consider popularity bias $\Z_j$. We postulate that the popularity bias of item $j$ is correlated with its number of interactions, traffic volume, and time; we use $\F_j$ to represent all aforementioned popularity factors and hope to learn the mapping function from $\F_j$ to $\Z_j$. We identify $\Z_j$ as a \emph{confounder}~\cite{pearl2009causality} which affects both exposed item $\V_j$ and observed interactions $\R_{ijk}$. 

Note that our \model can easily incorporate other types of in-domain bias which influence both exposed variables (users or items) and observed interactions, e.g. position bias ~\cite{joachims2017accurately}, by adding corresponding confounders in the graphical model similar to $\Z_j$. 
Here we explicitly know that $\Z_j$ models the popularity effect (since it is computed from the popularity factors); we therefore also call $\Z_j$ the \textit{explicit confounder}. This is in contrast with the cross-domain bias to be introduced next.

\textbf{Cross-domain Bias and Implicit Confounders.} 
Cross-domain bias such as changes in user community (US users' preferences may be very different from UK users') tends to be implicit or latent since relevant metadata is usually unavailable and therefore needs to be inferred from data. 


To address this challenge, \model defines the domain property $\D_k$ as an independent latent variable. $\D_k$ affects user behavioral pattern, item properties, and user-item interaction; we therefore posit treat it as a causal \emph{confounder} between user $\U_i$ and interaction $\R_{ijk}$ as well as between item $\V_j$ and interaction $\R_{ijk}$. Here we implicitly model the domain property $\D_k$ as a learnable latent variable which is not conditioned on any factors, we therefore call $\D_k$ the \textit{implicit confounder}.

\textbf{Distinguishing between In- and Cross-domain Biases.} 
Note that all users and items within each domain $k$ \emph{share} the same cross-domain bias induced by $\D_k$. In contrast, each item $j$ in a domain has \emph{different individual} in-domain bias induced by $\Z_j$. Therefore, during learning, \model will automatically extract the cross-domain bias shared across users and items into $\D_k$, while extracting the individual in-domain bias for each item $j$ into $\Z_j$.

\begin{figure}[t]
  \begin{center}
    \includegraphics[scale=0.46]{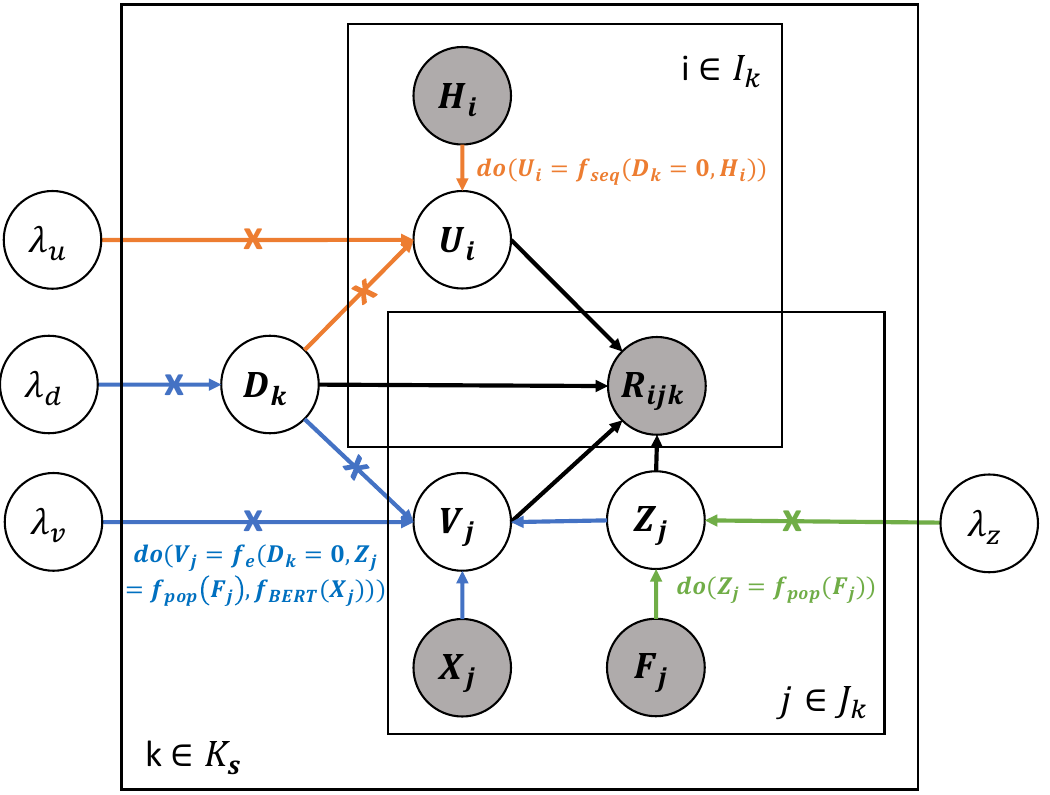}
  \end{center}
  \vskip -0.3cm
  \caption{For zero-shot recommendation in the target domain, we perform causal intervention on $\U_i$, $\V_j$, and $\Z_j$, i.e., $do\Big(\U_i=f_\textit{seq}(\D_k = \mathbf{0}, \H_i)\Big)$, $do\Big(\V_j=f_\textit{e}\big(\D_k = \mathbf{0}, \Z_j=f_\textit{pop}(\F_j), f_\textit{BERT}(\X_j)\big)\Big)$, and $do\Big(\Z_j=f_\textit{pop}(\F_j)\Big)$, to remove the cross-domain bias while injecting the in-domain bias in the target domain. }\label{fig:pgm-causal}
  \vskip -0.5cm
\end{figure}

\textbf{Pre-training.} For the pre-training stage of \model, we assume the user-item interactions $\R_{ijk}$ and item textual description $\X_j$ are observed for each domain, while the popularity factors $\F_j$ and user history $\H_i$ can be directly derived from the interaction $\R_{ijk}$. 
With hyperparameters $\lambda_u$, $\lambda_v$, $\lambda_d$, and $\lambda_z$ and given the graphical model ~\figref{fig:pgm}, the maximum a posteriori (MAP) estimation on latent variables ($\U_i,\V_j,\D_k,\Z_j$) can be decomposed as following:
\begin{gather}
    P(\U_i,\V_j,\D_k,\Z_j|\R_{ijk},\H_i,\X_j,\F_j,\lambda_{u},\lambda_{v},\lambda_{d},\lambda_{z}) \propto  \nonumber \\
    P(\R_{ijk}|\U_i,\V_j,\D_k,\Z_j) \cdot P(\U_{i}|\H_i,\D_k,\lambda_{u})\cdot P(\V_{j}|\X_j,\D_k,\Z_j,\lambda_{v}) \nonumber \\
    \cdot P(\Z_{j}|\F_{j}, \lambda_{z}) \cdot P(\D_k|\lambda_{d}). \label{eq:model_map}
\end{gather}

We define the conditional probability over the observed interactions as:
\begin{gather}
   P(\R_{ijk}|\U_i,\V_j,\D_k,\Z_j) = f_\textit{softmax}(\U_i^T \V_j + \D_k \W_d + \Z_j \W_z).
   \label{eq:rec_score}
\end{gather}
We assume Gaussian distributions on all latent variables in \figref{fig:pgm}, i.e., $\U_i$, $\V_j$, $\D_k$, and $\Z_j$, as follows:
\begin{align}
   P(\U_i|\H_i,\D_k,\lambda_u) &= \mathcal{N}(\U_i; f_\textit{seq}({\D_k, \H_i}), \lambda_u^{-1}\I_{B}), \nonumber \\
   P(\V_j|\X_j,\D_k,\Z_j,\lambda_v) &= \mathcal{N}(\V_j; f_\textit{item}(\D_k, \Z_j, f_\textit{e}(\X_j)), \lambda_v^{-1}\I_{B}),\nonumber  \\
   P(\Z_j|\F_j, \lambda_z) &= \mathcal{N}(\Z_j; f_\textit{pop}(\F_j), \lambda_z^{-1}\I_{B}),\nonumber \\
   P(\D_k|\lambda_d) &= \mathcal{N}(\D_k; \mathbf{0}, \lambda_d^{-1}\I_{B}),
   \label{eq:map_latent_variables}
\end{align}
where $\mathcal{N}(\x; \muu, \lambda^{-1}\I_{B})$ denotes the probability density function (PDF) of a Gaussian distribution with mean $\muu$ and diagonal covariance $\lambda^{-1}\I_{B}$ for the variable $\x$; $f_e(\cdot)$ and $f_{pop}(\cdot)$ are the learnable encoding functions for item embedding $\V_j$ and popularity embedding $\Z_j$, respectively. In our work, we choose to adopt multi-layer perceptron (MLP) for both $f_e(\cdot)$ and $f_{pop}(\cdot)$.

As in CDL~\cite{cdl} and ZESRec~\cite{ZESRec}, maximizing the posterior probability is equivalent to minimizing the negative log likelihood (NLL) of $\U_i, \V_j, \D_k$ and $\Z_j$ given $\R_{ijk}, \X_j, \H_i, \F_j, \lambda_u, \lambda_v, \lambda_d$ and $\lambda_z$:
\begin{gather}
   \mathcal{L} = \sum\nolimits_{k=1}^{K_s}\sum\nolimits_{i=1}^{I_k}\sum\nolimits_{j=1}^{J_k} -\log(f_\textit{softmax}(\U_i^T \V_j + \D_k \W_d + \Z_j \W_z)) \nonumber \\
   + \frac{\lambda_z}{2}\sum\nolimits_{j=1}^{J_k}||\Z_j - f_\textit{pop}(\F_j)||^2
   + \frac{\lambda_v}{2}\sum\nolimits_{k=1}^{K_s}\sum_{j=1}^{J_k}||\V_j - f_\textit{e}(\D_k, \Z_j)||^2  \nonumber\\
   + \frac{\lambda_u}{2}\sum\nolimits_{k=1}^{K_s}\sum\nolimits_{i=1}^{I_k}||\U_i - f_\textit{seq}({\D_k, \H_i})||^2
   + \frac{\lambda_d}{2}\sum\nolimits_{k=1}^{K_s}||\D_k||^2,
   \label{eq:map_loss}
\end{gather}
where $\W_d \in \mathbb{R}^B$, $\W_z \in \mathbb{R}^B$ are learnable vectors. {$K_s$ denotes a set of source domains; for domain $k$, we have $I_k$ users and $J_k$ items.}


\subsection{Zero-shot Recommendation} \label{subsec:zero_shot}
Next we discuss how to use the pre-trained model under the zero-shot setting as defined in~\cite{ZESRec}, which simulates an online environment where interaction data in the new domain is available only during inference, and all new users and items are unseen in source domains. 

\textbf{Causal Zero-shot Recommendation.} 
Given the pre-trained model, we enforce intervention on $\U_i$, $\V_j$ and $\Z_j$ by performing the \textit{do}-calculus~\cite{pearl2009causality} 
to eliminate the cross-domain bias while incorporate the in-domain bias in the target domain, {as shown in ~\figref{fig:pgm-causal}}. Specifically, we do not assume any domain properties and thus the posterior of $\D_k$ collapses to the prior $\mathbf{0}$; the popularity factors $\F_j$ are based on data statistics which can be derived on the fly in an online environment. Therefore, we compute $\U_i = f_\textit{seq}(\D_k = \mathbf{0}, \H_i)$ and $\V_j = f_\textit{e}(\D_k = \mathbf{0}, \Z_j=f_\textit{pop}(\F_j), f_\textit{BERT}(\X_j))$. Note that here we set $\D_k=\mathbf{0}$ for $\U_i$ and $\V_j$. This serves as an approximation to the output expectation over the distribution of $\D_k$ as input, and we found this approach achieved similar performance in practice. Following the \textit{back-door formula}~\cite{pearl2009causality} we have: 
\begin{align}
   &P(\R_{ijk}|{do}(\U_i, \V_j, \Z_j)))  \nonumber\\
   &= \int P(\R_{ijk}|\U_i,\V_j,\D_k,\Z_j) P(\D_k) d\D_k \nonumber\\
   &= \int f_\textit{softmax}(\U_i^T \V_j + \D_k \W_d + \Z_j \W_z) P(\D_k) d\D_k \nonumber\\
   &= \int \frac{\textit{exp}(\U_i^T \V_j + \Z_j \W_z) \textit{exp}(\D_k \W_d)}{\sum_j \textit{exp}(\U_i^T \V_j + \Z_j \W_z)\textit{exp}(\D_k \W_d)}  P(\D_k)d\D_k \nonumber\\
   &= \int \frac{\textit{exp}(\U_i^T \V_j + \Z_j \W_z)}{\sum_j \textit{exp}(\U_i^T \V_j + \Z_j \W_z)}  P(\D_k)d\D_k \nonumber\\
   &= f_\textit{softmax}(\U_i^T \V_j + \Z_j \W_z). \label{eq:zeroshot_inf}
\end{align}
Note that for $f_\textit{softmax}(\cdot)$ we only consider items in the same domain.




\subsection{Fine-tuning}
Once the interaction data in the target domain is available, we can then fine-tune all the parameters of \model end-to-end in this new domain. During fine-tuning, we optimize \eqnref{eq:map_loss} on the target domain data and re-estimate all latent variables along with corresponding encoding functions as stated in \eqnref{eq:rec_score} and \eqnref{eq:map_latent_variables}. 

For inference after fine-tuning, with the re-estimated latent vectors $(\hat{\D}_k, \hat{\Z}_j, \hat{\V}_j, \hat{\U}_i)$ and all learnable parameters inside $f_\textit{softmax}(\cdot)$, i.e., $\hat{\W}_d$ and $\hat{\W}_z$, we then perform causal inference by intervening on the domain bias ($\D_k$), popularity bias ($\Z_j$), item properties ($\V_j$), and user interests ($\U_i$): 
\begin{gather*}
   P(\R_{ijk}|{do}(\U_i, \V_j, \D_k, \Z_j)) = f_\textit{softmax}(\hat{\U}_i^T \hat{\V}_j + \hat{\D}_k \hat{\W}_d + \hat{\Z}_j \hat{\W}_z).
\end{gather*}

\section{Experiments} \label{sec:experiment}

In this section, we first introduce the experiment set up, and then present the results and analysis with major goals to address the following questions:
\begin{itemize}[leftmargin=17pt]
    \item[{\bf Q1}] When training on multiple domains and testing on a new domain, will \model bring benefits to the new domain? What's the \emph{zero-shot} performance of \model compared with baselines?
    \item[{\bf Q2}] {How effective is the proposed casual debiasing mechanism to alleviate in- and cross-domain biases? To what extent does it improve the performance?}
    \item[{\bf Q3}] Given the pre-trained model, how does \model perform compared to baselines {if \emph{full fine-tuning}} (more details in~\secref{sec:Fine-tune}) in the new domain is allowed? 
    \item[{\bf Q4}] Given the pre-trained model, if fine-tuning in the new domain is allowed, how does the \emph{number of fine-tuning samples} affect final performance?
\end{itemize}

\subsection{Dataset Processing and Statistics}
\label{sec:Datasets}
We consider two datasets: (1) \textbf{XMarket} dataset\footnote{https://xmrec.github.io/}~\cite{bonab2021crossmarket}, which is a large-scale real-world dataset covering 18 local markets (countries) on 16 different product categories; and (2) \textbf{Online Retail}\footnote{https://www.kaggle.com/datasets/carrie1/ecommerce-data}, which contains data collected from a UK-based online retail platform.

To ensure a sufficient number of non-overlapping users/items to evaluate zero-shot and fine-tuning performance rigorously while guaranteeing enough user-item interactions for training and evaluation, we set \textit{``India'', ``Spain'', ``Canada''} in the XMarket dataset as the \textbf{pre-trained datasets}, set \textit{``Australia'', ``Mexico'', ``Germany'', ``Japan''} in the XMarket dataset as the \textbf{cross-market datasets}, and choose the online retail dataset as the \textbf{cross-platform dataset}. 
For experiments, we pre-train \model on the pre-trained datasets and evaluate it on both cross-market datasets and cross-platform datasets.

\textbf{Domain and Dataset Split.} 
{For each domain/dataset, we randomly split users into training/validation/test sets with the ratio 4:3:3.
The test splits of target domains are used for evaluation under both the zero-shot setting and the fine-tuning setting. 
Furthermore, to evaluate the zero-shot performance rigorously, we further filter out the overlapping users and items from the test splits; we name the remaining split \textbf{unseen} test set.
{See the supplement for the dataset statistics of each target domain. }


\subsection{Evaluated Methods}
\label{sec:Compared Methods}
We compared our proposed \model with state-of-the-art methods, including \textbf{Random} (recommend by random selection), \textbf{POP} (recommend by popularity), \textbf{\sbert}~\cite{reimers-2019-sentence-bert}, \textbf{\gruS}~\cite{gru4rec}, \textbf{\transformerS}~\cite{sasrec}, \textbf{\gru}~\cite{ZESRec}, \textbf{\unisrec}~\cite{abs-2206-05941}, and \textbf{\transformer}. Please see the Appendix~\ref{sec:Modeling/Training Details} for implementation details.

\begin{table*}[t]
\vskip -0.0cm
\centering
\begin{scriptsize}
\resizebox{0.95\textwidth}{!}{
\begin{tabular}{ccccccccccc}
\toprule
Scenario                                                                   & Dataset                                                                  & Metric                      & Test Type & Random &    POP       & SBERT               & \gru & \unisrec & \transformer             & \model \\ \midrule
\multirow{16}{*}{\begin{tabular}[c]{@{}c@{}}Cross\\ -Market\end{tabular}}  & \multirow{4}{*}{Australia}                                               & \multirow{2}{*}{Recall@K\%} & all       & 0.0004 & 0.0450 & 0.0552             & 0.0431 & 0.0404  & \underline{0.0583}     & \textbf{0.1036} \\
                                                                          &                                                                          &                             & unseen    & 0.0004 & 0.0127 & 0.0619              & 0.0472 & 0.0452  & \underline{0.0655}     & \textbf{0.0756} \\
                                                                          &                                                                          & \multirow{2}{*}{r-NDCG@K\%} & all       & 0.0002 & 0.0261 & 0.0380              & 0.0287 & 0.0281  & \underline{0.0396}     & \textbf{0.0656} \\
                                                                          &                                                                          &                             & unseen    & 0.0002 & 0.0063 & 0.0420              & 0.0323 & 0.0316  & \underline{0.0438}     & \textbf{0.0476} \\ \cmidrule{2-11}
                                                                          & \multirow{4}{*}{Mexico}                                                  & \multirow{2}{*}{Recall@K\%} & all       & 0.0004 & 0.0695 & 0.1509              & 0.1397 & 0.1521  & \underline{0.1645}     & \textbf{0.2316} \\
                                                                          &                                                                          &                             & unseen    & 0.0004 & 0.0248 & 0.1403              & 0.1157 & 0.1386  & \textbf{0.1487}        & \underline{0.1469} \\
                                                                          &                                                                          & \multirow{2}{*}{r-NDCG@K\%} & all       & 0.0002 & 0.0388 & 0.1095              & 0.0961 & 0.1134  & \underline{0.1178}     & \textbf{0.1557} \\
                                                                          &                                                                          &                             & unseen    & 0.0002 & 0.0111 & 0.1009              & 0.0748 & 0.1000  & \textbf{0.1027}        & \underline{0.1017} \\ \cmidrule{2-11}
                                                                          & \multirow{4}{*}{Germany}                                                 & \multirow{2}{*}{Recall@K\%} & all       & 0.0004 & 0.1514 & 0.2737              & 0.2639 & 0.2703  & \underline{0.2827}     & \textbf{0.3526} \\
                                                                          &                                                                          &                             & unseen    & 0.0004 & 0.1016 & \underline{0.2528}  & 0.2295 & 0.2430  & 0.2506                 & \textbf{0.2750} \\
                                                                          &                                                                          & \multirow{2}{*}{r-NDCG@K\%} & all       & 0.0002 & 0.0936 & \underline{0.2103}  & 0.1886 & 0.2100  & \underline{0.2103}     & \textbf{0.2576} \\
                                                                          &                                                                          &                             & unseen    & 0.0002 & 0.0559 & 0.1910              & 0.1703 & 0.1852  & \underline{0.1923}     & \textbf{0.1954} \\ \cmidrule{2-11}
                                                                          & \multirow{4}{*}{Japan}                                                   & \multirow{2}{*}{Recall@K\%} & all       & 0.0004 & 0.0657 & \underline{0.2926}  & 0.2259 & 0.2579  & 0.2817                 & \textbf{0.3083} \\
                                                                          &                                                                          &                             & unseen    & 0.0004 & 0.0556 & \underline{0.2737}  & 0.2152 & 0.2662  & 0.2730                 & \textbf{0.2741} \\
                                                                          &                                                                          & \multirow{2}{*}{r-NDCG@K\%} & all       & 0.0002 & 0.0376 & \underline{0.2016}  & 0.1514 & 0.1776  & 0.1929                 & \textbf{0.2033} \\
                                                                          &                                                                          &                             & unseen    & 0.0002 & 0.0325 & \textbf{0.2061}     & 0.1545 & 0.1967  & 0.1949                 & \underline{0.1980} \\ \midrule
\multirow{2}{*}{\begin{tabular}[c]{@{}c@{}}Cross\\ -Platform\end{tabular}} & \multirow{2}{*}{\begin{tabular}[c]{@{}c@{}}Online\\ Retail\end{tabular}} & \multirow{1}{*}{Recall@K\%} & all/unseen   & 0.0050 & 0.0672 & 0.1383          & 0.0564 & 0.1147  & \underline{0.1397}                 & \textbf{0.1794}  \\
                                                                          &                                                                          & \multirow{1}{*}{r-NDCG@K\%} & all/unseen    & 0.0025 & 0.0362 & \underline{0.0848}          & 0.0316 & 0.0698  & 0.0841                 & \textbf{0.1065} \\ \bottomrule
\end{tabular}}
\caption{Zero-shot performance comparison of different methods. K\%$=0.04\%$ for Cross-Market and K\%$=0.5\%$ for Cross-Platform. We mark the best results with \textbf{bold face} and the second best results with \underline{underline}.}\label{table:zero-shot}
\end{scriptsize}
\vskip -0.6cm
\end{table*}

\begin{table*}[]
\centering
\begin{scriptsize}
\resizebox{0.9\textwidth}{!}{
\begin{tabular}{cccccccccc}
\toprule
Scenario                                                                   & Dataset                                                                  & Metric     & Random & POP    & SBERT  & $\text{\gruS}^*$ & $\text{\transformerS}^*$ & \unisrec & \model \\ \midrule
\multirow{8}{*}{\begin{tabular}[c]{@{}c@{}}Cross\\ -Market\end{tabular}}   & \multirow{2}{*}{Australia}                                               & Recall@K\% & 0.0004 & 0.0450 & 0.0552 & 0.0546          & \underline{0.0735}  & 0.0600              & \textbf{0.1130} \\
                                                                          &                                                                          & r-NDCG@K\% & 0.0002 & 0.0261 & 0.0380 & 0.0349          & \underline{0.0495}  & 0.0416              & \textbf{0.0715} \\ \cmidrule{2-10}
                                                                          & \multirow{2}{*}{Mexico}                                                  & Recall@K\% & 0.0004 & 0.0695 & 0.1509 & 0.2315          & 0.2475              & \underline{0.2478}  & \textbf{0.2646} \\
                                                                          &                                                                          & r-NDCG@K\% & 0.0002 & 0.0388 & 0.1095 & 0.1683          & 0.1813              & \textbf{0.1832}     & \underline{0.1825} \\ \cmidrule{2-10}
                                                                          & \multirow{2}{*}{Germany}                                                 & Recall@K\% & 0.0004 & 0.1514 & 0.2737 & 0.3344          & \underline{0.3428}  & 0.3235              & \textbf{0.3935} \\
                                                                          &                                                                          & r-NDCG@K\% & 0.0002 & 0.0936 & 0.2103 & 0.2653          & \underline{0.2726}  & 0.2697              & \textbf{0.2964} \\ \cmidrule{2-10}
                                                                          & \multirow{2}{*}{Japan}                                                   & Recall@K\% & 0.0004 & 0.0657 & 0.2926 & 0.3428          & \underline{0.3645}  & 0.3587              & \textbf{0.3813} \\
                                                                          &                                                                          & r-NDCG@K\% & 0.0002 & 0.0376 & 0.2016 & 0.2475          & \underline{0.2554}  & 0.2549              & \textbf{0.2601} \\ \midrule
\multirow{2}{*}{\begin{tabular}[c]{@{}c@{}}Cross\\ -Platform\end{tabular}} & \multirow{2}{*}{\begin{tabular}[c]{@{}c@{}}Online\\ Retail\end{tabular}} & Recall@K\% & 0.0050 & 0.0672 & 0.1383 & 0.2524          & \underline{0.2728}              & 0.2513              & \textbf{0.2992}       \\
                                                                          &                                                                          & r-NDCG@K\% & 0.0025 & 0.0362 & 0.0848 & 0.1545          & \underline{0.1665}              & 0.1589              & \textbf{0.1769}       \\ \bottomrule
\end{tabular}}
\caption{Full fine-tuning results of different models. K\%$=0.04\%$ for Cross-Market and K\%$=0.5\%$ for Cross-Platform. The notation $^*$ indicates the model is trained from scratch. We mark the best results with \textbf{bold face} and the second best results with \underline{underline}.}\label{table:fine-tiune}
\end{scriptsize}
\vskip -0.8cm

\end{table*}

\subsection{Evaluation Metrics}
\label{sec:Metric}
We use Recall@K\% and our proposed r-NDCG@K\% to evaluate model performance.
For pre-training on multiple source domains, it is not ideal to evaluate performance and apply early stopping based on averaged Recall@K or NDCG@K.
This is because different domains have different numbers of items, and domains with fewer items tend to have larger NDCG@K or Recall@K, consequently dominating the evaluation.
Therefore, we use Recall@K\% and our proposed r-NDCG@K\% (`r' stands for `revised') to do model evaluation and early-stop.
Denoting the number of items in a domain as $N$, with the next interacted item as the single target, r-NDCG@K\% is defined as follows:
\begin{gather}
   \text{r-DCG} = 1/\log(a+\frac{b*(rank-1)}{N}),
   \label{eq:NDCG0}
\end{gather}
\begin{gather}
   \text{r-iDCG} = 1/\log(a+\frac{b*(1-1)}{N}) = 1/\log(a),
   \label{eq:NDCG1}
\end{gather}
\begin{gather}
   \text{r-NDCG} = \frac{\text{r-DCG}}{\text{r-iDCG}} = \frac{\log(a)}{\log(a+\frac{b*(rank-1)}{N})}.
   \label{eq:NDCG2}
\end{gather}
Here $a$ and $b$ are hyperparameters, 
and $N$ is the number of items in a given domain.
With this approach, the value of r-NDCG@K\% is normalized by the number of items $N$ in that domain.
As a result, the value will not tend to be high when $N$ is low.
This a situation where the r-NDCG@K\% of domain with small $N$ dominates r-NDCG@K\% of domain with large $N$.

Note that r-NDCG becomes the commonly used NDCG if $a=2$ and $b=N$. In our experiments, we set $a=2$ and $b=15000$. Meanwhile, we use @K\% instead of @K, with @K\% indicating that only the top K\% recommended items will contribute to the Recall@K\% and r-NDCG@K\%. Empirically, we set K\%$=$0.04\% for the cross-market scenario and set K\%$=$0.5\% for the cross-platform scenario. 
We use the average r-NDCG@K\% on the validation set to early-stop training; we use both r-NDCG@K\% and Recall@K\% on the test set to evaluate a model.

\begin{figure*}[t]
    \centering
    \includegraphics[width=1.00\textwidth]{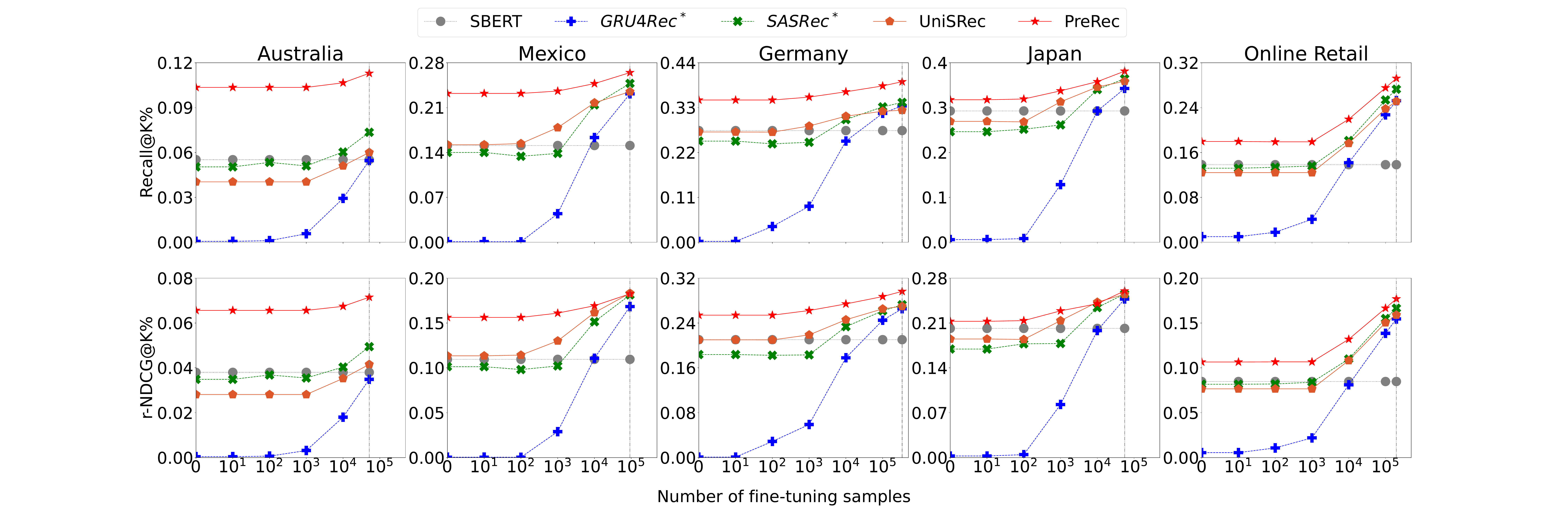}
    \vskip -0.4cm
    \caption{Incremental fine-tuning results of different models. K\%$=0.04\%$ for Cross-Market and K\%$=0.5\%$ for Cross-Platform. The solid line indicates the model is pre-trained while the dashed line indicates the model is trained from scratch. The last point for each line corresponds to the full fine-tuning with all available training data in the target domain.}
    \vskip -0.2cm
    \label{fig:incremental}
\end{figure*}

\subsection{Zero-shot Experiments}
\label{sec:Zero-shot}
{For the zero-shot experiment, we pre-train \model on three source domains (\textit{``India''}, \textit{``Spain''}, and \textit{``Canada''} in the XMarket dataset) and evaluate its performance against a variety of baselines on five target domains (\textit{``Australia''}, \textit{``Mexico''}, \textit{``Germany''}, \textit{``Japan''} in the XMarket dataset and the Online Retail dataset) respectively. }

{\textbf{Evaluation Setup.} During evaluation, we simulate an online environment where models \textit{can only access the target domain data during inference}, and all the models are evaluated on the test set of the target domain. 
Besides measuring performance on the whole test set, we also separately measure performance on the \emph{unseen} test set (see~\secref{sec:Datasets}). 
We utilize Recall@K\% and r-NDCG@K\% as evaluation metrics since they are less sensitive to differences in task difficulty across different domains. This is to ensure that performance for different domains is directly comparable, which is reflected by the consistent performance of the Random model across four target domains within the XMarket dataset (Recall@K\% and r-NDCG@K\% are constantly 0.004 and 0.002, respectively). The baselines can be broadly categorized into (1) \textbf{non-learnable zero-shot models}: Random, POP, SBERT, and (2) \textbf{learnable zero-shot models}: ZESRec, UnisRec, \transformer. Table~\ref{table:zero-shot} shows the zero-shot performance of different models.
}

\textbf{Zero-shot Results (Q1).}
For the baselines, among the non-learnable zero-shot models, SBERT significantly outperforms both Random and POP since it leverages additional information-rich item descriptions. Importantly, SBERT achieves comparable performance with the learnable zero-shot models, which shows that without proper pre-training strategy to filter out the noise in the training data, the pre-training stage may not bring large performance gain due to overfitting on source domains. Among the learnable zero-shot models, the UnisRec slightly outperforms the ZESRec because it includes additional contrastive training objectives and uses a self-attention-based sequential model instead of GRU in ZESRec. 
Our proposed \model significantly outperforms all baselines by a large margin on almost all cases under both cross-market and cross-platform scenarios thanks to the causal debiasing mechanism which extracts the generic knowledge across source domains while reducing biases in data. 

The improvement on unseen items is relatively lower than the overall improvement. We discovered that unseen items are usually unpopular items in target domains; this is also why POP's performance is worse on ``unseen'' items than in ``all'' items. 

{For the zero-shot results of \model among cross-market datasets, we observed the largest improvement on Australia, moderate improvement on Germany and Mexico, with the least improvement on Japan. Such differences in improvements are potentially due to the differences in similarity between source domains and target domains: users in Australia share more common interests with users in Canada, which is the largest source domain; German speaks the same language as Canadian and Indian, and so does Mexican and Spanish; none of the source domains are Japanese-speaking countries, and Japanese users' interests tend to be different from those in the source domains due to differences in culture.}

\textbf{Ablation Study on Causal Debiasing Mechanism (Q2).}
{To verify the effectiveness of the causal debiasing mechanism, we performed ablation studies on all four target domains. We implemented \transformer, a simplified version of \model, which ignores the cross-domain and in-domain bias terms during pre-training. Table~\ref{table:zero-shot} shows that in almost all cases \model outperforms \transformer.}

\subsection{Fine-tuning Experiments}
\label{sec:Fine-tune}
{For the fine-tuning experiments, we test under} two settings: (1) \emph{full fine-tuning} where we train or fine-tune models on all available training data in the target domain, {(i.e., the training split of the target domain)}, and (2) \emph{incremental fine-tuning} where we train or fine-tune models on a set of incrementally larger training data in the target domain. {All the models are evaluated on the test set of the target domain. There are three sets of baselines: (1) \textbf{non-learnable zero-shot model}: Random, POP, SBERT; (2) \textbf{In-domain model}: GRU4Rec, SASRec; (3) \textbf{Pre-trained model}: UniSRec.}

\textbf{Full Fine-tuning (Q3).} 
Table~\ref{table:fine-tiune} shows the full fine-tuning performance of different methods. {All in-domain models (GRU4Rec and SASRec) achieve better performance than SBERT, which is the best non-learnable zero-shot model
. The pre-trained model UniSRec achieves performance similar to the in-domain models', showing that UniSRec brings no additional gain from pre-training on other domains if there is sufficient data in target domains. \model outperforms all baselines on almost all cases under both cross-market and cross-platform scenarios by a large margin after fine-tuning, demonstrating that: (1) \model distills 
generic knowledge in the pre-training stage, and it still has complementary value even after fine-tuning on target domain data; (2) \model is capable of rapidly adapting to the new domain without forgetting the distilled knowledge during pre-training (i.e., robust to catastrophic forgetting).}

\textbf{Incremental Fine-tuning (Q4).} 
Fig~\ref{fig:incremental} shows the results for incremental fine-tuning. Among all target domains, \model 
outperforms all baselines, with the performance steadily improving as the number of target domain training samples increases, showing \model is progressively adapting to the target domain; the performance gap between \model and the baselines is gradually shrinking, suggesting diminishing complementary value of the distilled knowledge during pre-training. It is worth noting that the performance gap between \model and baselines is still prominent even when there are as many as $10^4$ target domain training samples, 
equivalent to several weeks/months of effort on collecting data.
\vspace{-0.25 cm}
\section{Related Work}
\textbf{In-domain Recommendation.} 
There is a rich literature on in-domain recommendation, i.e., training and testing the recommender systems on the same domain. Collaborative filtering methods such as PMF~\cite{pmf} and BPR~\cite{bpr} are first proposed to approach this problem. Later the deep learning methods such as GRU~\cite{ChungGCB14}, Transformer~\cite{vaswani2017attention} and Graph Neural Network~\cite{srgnn} were proposed and demonstrated superiority on a variety of tasks, works such as GRU4Rec~\cite{gru4rec}, SAS4Rec~\cite{sasrec} and KGAT~\cite{wang2019kgat} adopt these latest neural network architectures and achieve great success in the recommendation regime. These methods assume only rating/interaction data is available without any content information (e.g., text description for items). Such an assumption precluded their direct application to pretraining-based recommenders. A new line of deep learning recommenders, pioneered by collaborative deep learning (CDL)~\cite{cdl} and its variants~\cite{CRAE,ColVAE}, seamlessly incoporate content information into deep recommenders, thereby opening up the possibility of pretraining-based recommenders and significantly alleviating cold-start problems in recommender systems. 

\textbf{Cross-domain Recommendation.} 
Beyond in-domain recommendation, there are other works that build on domain adaptation methods~\cite{VDI,TSDA,UDIL,GRDA,CIDA,DANN} from the machine learning community to achieve cross-domain recommendation. Specifically, they utilize data from source domains to boost recommendation performance in target domains with either common users or common items~\cite{DBLP:conf/sigir/Yuan0KZ20,DBLP:conf/sigir/WuYCLH020,DBLP:conf/sigir/BiSYWWX20a, li2019zero, DBLP:conf/sigir/Hansen0SAL20,DBLP:conf/sigir/LiangXYY20,DBLP:conf/sigir/ZhuSSC20, liu2020heterogeneous}. Another line of work, usually referred to as Dual-Target Cross-Domain Recommendation, tries to improve performance of both source and target domains~\cite{0008T20,HuZY18,ZhaoLF19,LiuLLP20} (a more comprehensive survey of cross-domain recommender systems can be found in~\cite{ZhuW00L021}). Note that these aforementioned cross-domain recommendation methods are fundamentally different from pre-trained recommendation. 

\textbf{Bias in Recommendation.} 
While recommender systems are analyzed through in-domain and cross-domain perspectives, there is
an emphasis on understanding them through the lens of bias~\cite{chen2023bias}.
Within a single domain, the topic of item popularity debiasing has garnered significant attention.
Studies~\cite{ZhangF0WSL021, MaNLD20, abs-2006-11011} have demonstrated that addressing item popularity bias can enhance recommendation quality.
In the broader context of cross-domain bias mitigation, 
~\cite{LiYMZLGDW21} has adapted traditional Inverse-Propensity-Score (IPS) to fit cross-domain contexts, while~\cite{DuWF0022} employs invariant representations to reduce the impact of spurious correlations.
However, the method to effectively leverage item popularity knowledge from pre-training domains to new domains remains an open question.

\textbf{Causality in Recommendation.} 
Closely related to debiasing are studies on causality-inspired recommender systems. 
~\cite{wang2020causal,Wang0LZY022} extract invariant user-item embeddings across varying scenarios, aiming to minimize the influence of potentially misleading features.
Both~\cite{zhang2021causal} and~\cite{zheng2021disentangling} integrate the influence of popularity into their causal analyses within a specific domain. 
Nonetheless, the application of causality on a broader scale, specifically within pre-trained recommender systems, remains largely unexplored.

\vspace{-0.35 cm}
\section{Conclusion}
In this paper, we identify two types of bias: (1) in-domain bias: we consider the item popularity bias taking effects within each domain; and (2) cross-domain bias: we consider variability introduced by the unique domain properties. We propose \model equipped with a novel causal debiasing mechanism to deal with both bias terms. We extensively conduct experiments under both cross-market and cross-platform scenarios and demonstrate the effectiveness of our model. Future work includes identifying other confounders to incorporate into our causal debiasing framework, experimenting with better PLMs such as OPT~\cite{opt}, exploring other modalities, etc.

\bibliographystyle{ACM-Reference-Format}
\bibliography{main0WSDM0PreRec}

\blankpage
\appendix
\begin{center}
      {\Large \bf {Pre-trained Recommender Systems: A Causal Debiasing
Perspective (Supplementary Material)} \par}
     
      \vskip .5em
      \vspace*{12pt}
\end{center}

\begin{figure*}[t]
    \centering
    \includegraphics[width=0.90\textwidth]{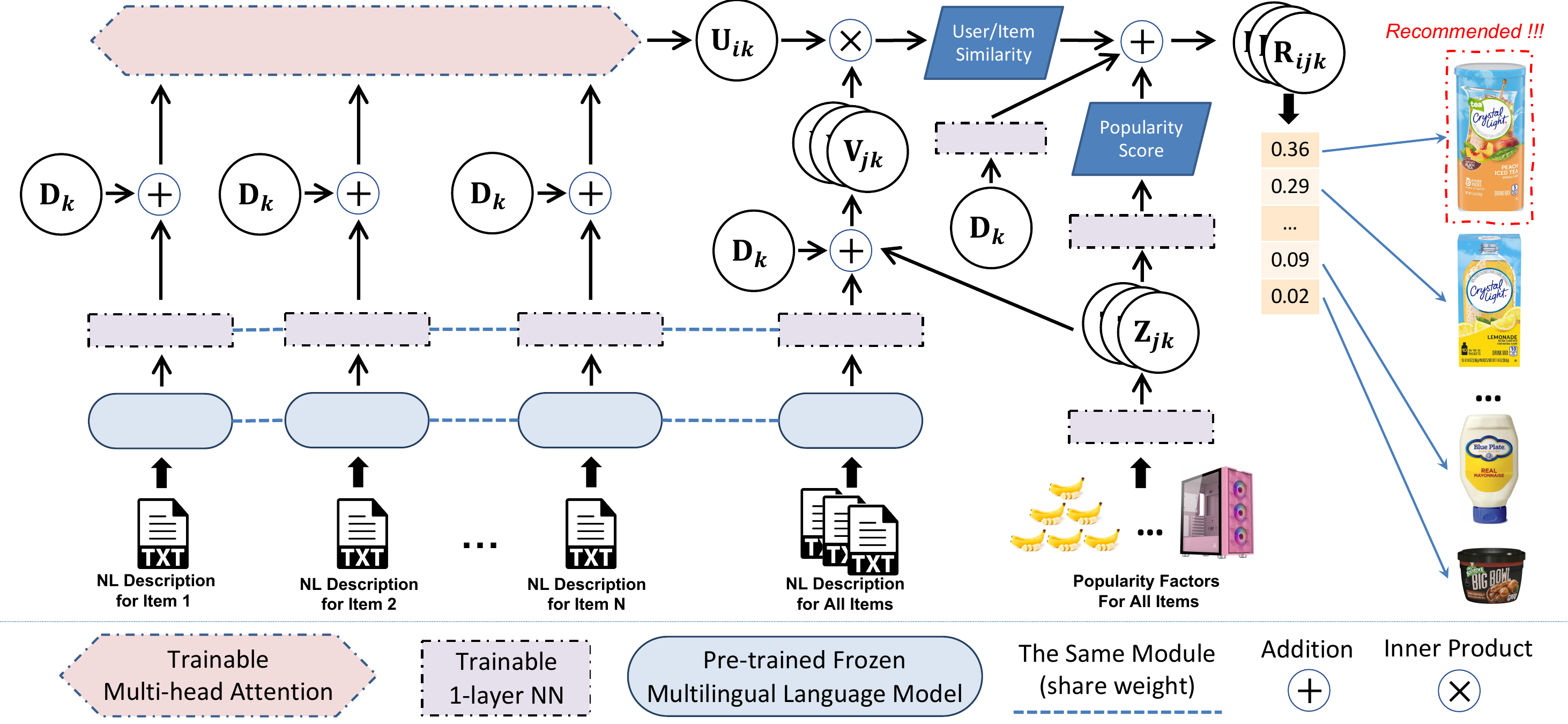}
    \caption{The neural network implementation for our \model. Analogous to Fig. 2 in the main paper, $f_{\text{BERT}}$ is implemented as a multilingual language model, $f_{\text{seq}}$ is implemented as a transformer decoder, and $f_{\text{pop}}$ is implemented as a linear layer with activation.}
    \label{fig:nn}
\end{figure*}

\begin{figure*}[t]
    \centering
    \includegraphics[width=0.80\textwidth]{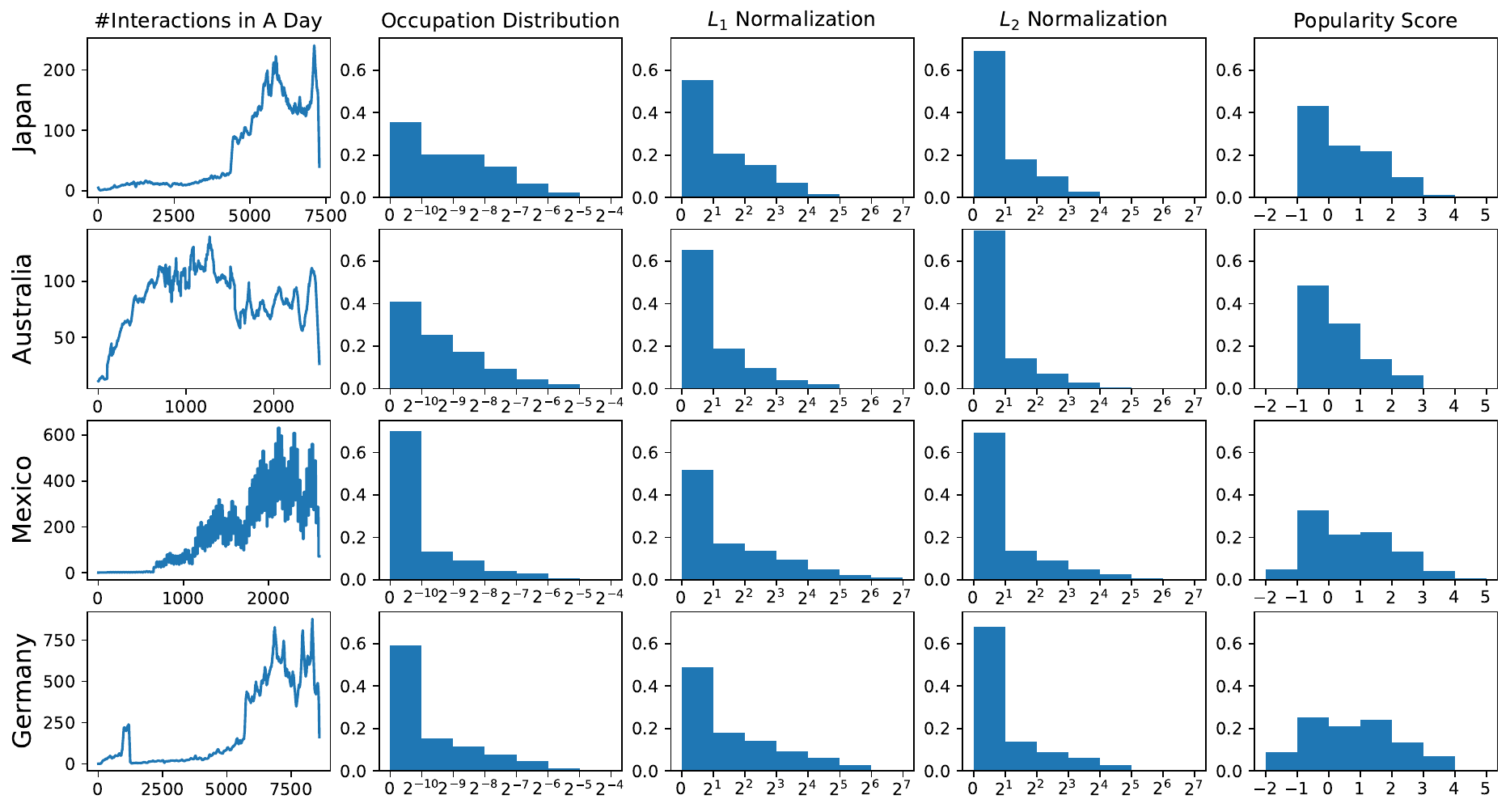}
    \vskip -0.5cm
    \caption{The illustration of how popularity score works. The first column shows different domains have different traffic volumes and the second column shows occupation is not comparable among different domains since different domains have different numbers of available items. The third and fourth columns show the distribution of the first and the second dimension of the popularity factor $\F$ in each domain, which is a normalized frequency proposed in this paper, and shown to be more comparable among different domains. After pertaining, the popularity factor $\F$ is further mapped to the popularity score. As shown in the last column, the popularity score is comparable among zero-shot domains, indicating the generalizability of our proposed method for popularity.}
    \vskip -0.3cm
    \label{fig:pop}
\end{figure*}

\section{Neural Network Architecture}
Figure~\ref{fig:nn} shows the neural network architecture of our \model.
The item overall property $\V_{jk}$ is a summation of domain property $\D_k$, item popularity property $\Z_{jk}$, and item embedding extracted from item description via a pre-trained frozen multilingual language model and a linear layer.
A sequential model will aggregate item embeddings of items in user history to generate user embedding $U_{ik}$ via multi-head attention.
Then user/item similarity is the inner product between $\U_{ik}$ and $\V_{jk}$.
This similarity will be further calibrated by popularity score and domain bias to form the use/item preference $\R_{ijk}$.

\section{Implementation Details}
\label{sec:Modeling/Training Details}
Here we note down the details for model implementations and the training process.

We consider the following methods during evaluation:
\begin{itemize}[leftmargin=15pt]
    \item \textbf{Random}: Recommend items by random selection from the whole item catalogue without replacement.
    \item \textbf{POP}: Recommend items based on the items' popularity in the last stage (i.e., the last 15 days in our experiments). 
    \item \textbf{\sbert}: Apply the pre-trained SBERT~\cite{reimers-2019-sentence-bert} on the item's description and generate the textual embedding as the item embedding, use the item embedding of the last interacted item in the user sequence as the user embedding, and recommend the next item to the user based on the inner product between the user embedding and the item embedding. 
    \item \textbf{\gruS}~\cite{gru4rec}: Use GRU to model the user interaction sequence for session-based recommendation. For fair comparison, we use the SBERT embedding generated from item's description as the input item representation.
    \item \textbf{\transformerS}~\cite{sasrec}: A self-attention based sequential model for session-based recommendation. For fair comparison, we use the SBERT embedding generated from item's description as the input item representation.
    \item \textbf{\gru}~\cite{ZESRec}: A flexible sequential framework that can be pre-trained on a source domain and directly applied to a target domain. It generates the universal item embedding via pre-trained language models (in our experiments we adopt SBERT) and generates universal user embeddings by aggregating universal item embeddings in the user sequence. In the experiments, we employ the GRU variant of ZESRec.
    \item \textbf{\unisrec}~\cite{abs-2206-05941}: A follow-up work of ZESRec which further enables pre-training on multiple source domains. It adopts a self-attention based sequential model, equipping it with a MoE-enhanced Adaptor and additional contrastive training objectives, to assist with domain fusion and improve performance. {As \cite{abs-2206-05941} claims, UniSRec is a state-of-the-art cross-domain model.}
    \item \textbf{\transformer}: A simplified version of \model without the causal debiasing mechanism; in other words it does not take the cross-domain bias and the in-domain bias into the consideration during pre-training.
\end{itemize}


\textbf{Models.} 
All methods except Random and POP use Sentence-BERT~\cite{reimers-2019-sentence-bert} to extract item BERT embeddings. For a fair comparison, all methods use $256$-dimensional item embeddings and user embeddings; {for sequential models based on GRU, the number of layers of GRU is set to 2; for self-attention based sequential models, the number of multi-head attention layers is set to 2.}

\textbf{Training.} 
We use Adam~\cite{KingmaB14} for both pre-training and fine-tuning with the learning rate of 0.0003. 
L2 regularization is implemented with weight decay in PyTorch's Adam. 
We use the average r-NDCG@K\% on the validation set to early-stop pre-training, after which the pre-trained model is used for fine-tuning on each target domain. See the supplement for more details.

\section{The Design Rationale of Computing Popularity
Properties}
\label{app:pop}
To calculate the popularity score in a domain $k$, assuming a set of items $J_k$, and the number of interactions of an item $j$ in time slot $T$ is $c_j^T$. Naively, one could compute popularity factors $\F_j$ via the following two options:
\begin{compactitem}
\item \textbf{Method A:} Directly using $\log{c_j^T}$ or $c_j^T$ as the popularity factors $\F_j$ for item $j$ at time interval $T$.
\item \textbf{Method B:} Computing the percentage of the traffic volume, i.e., using $c_j^T/\sum_{i\in J_k} c_i^T$ or $(c_j^T/\sum_{i\in J_k} c_i^T)^p$ as the popularity factors $\F_j$ for item $j$ at time interval $t$, where $p$ is a pre-defined scalar. 
\end{compactitem}

However, these two methods are not transferable across multiple domains:
\begin{compactitem}

\item For \textbf{Method A}, assuming domain $m$ has more interactions than domain $n$ and both domains have the same number of items. In this case, even though item $j$ in domain $m$ has more interactions than item $i$ in domain $n$, it does not mean that item $j$ in domain $m$ is more popular than item $i$ in domain $n$. 
\item For \textbf{Method B}, assuming domain $m$ has more items than domain $n$. In this case, even if item $j$ in domain $m$ has the same percentage of traffic volume (e.g. 5\%) as item $i$ in domain $n$, item $j$ may still be more popular in domain $m$ than item $i$ in domain $n$.
\end{compactitem}

To address the abovementioned issue, we design a method inspired by the half Gaussian distribution. Assuming in any domain, the number of interactions of each item follows a half Gaussian distribution. The popularity factor $\F_j$ of item $j$ can be measured as:
\begin{align}
    \F_j = c_j^T/\sqrt{\frac{\sum_{i\in J_k} (c_i^T)^2}{|J_k|}},\label{pop1}
\end{align}
where the popularity factor $\F_j$ is transferable from domain to domain. 

We check the distribution over the number of interactions of items for each real-world dataset and discover that they may not precisely follow the half Gaussian distribution. Therefore, we propose to consider different orders of norms for the denominator in ~\eqnref{pop1} and compute a set of popularity factors as $\F_j$:
\begin{gather*}
   \F_j = [\frac{c_j^{T}}{s_1^{T}}, \frac{c_j^{T}}{s_2^{T}}, ... ,\frac{c_j^{T}}{s_w^{T}}],
\end{gather*}
where $s_w^{T}$ is a normalization term calculated as:
\begingroup\makeatletter\def\f@size{8}\check@mathfonts
\begin{gather*}
   s_w^{T} = (\sum_{j\in J_k} (c_j^{T})^w / |J_k|)^{\frac{1}{w}}
\end{gather*} 
with varied $k$ as the popularity property, and use a trainable neural network to get the popularity score. The final $w$ ranges from 1 to 4, i.e., the popularity property factors $\F_j$ is a 4-dimension vector.

~\figref{fig:pop} illustrates how the popularity score works on four zero-shot domains.



\section{The Choice of K\% in r-NDCG@K\% and Recall@K\%}
We choose $K\%=0.04\%$ due to under the number of samples $N$ from Table 3 we have $10\leq N \times K\%\leq50$. This provides the results between NDCG@10 and NDCG@50, which aligns with common evaluation setting including NDCG@10, NDCG@20, or NDCG@50.

\begin{table*}
\centering
\resizebox{0.75\textwidth}{!}{
\begin{tabular}{cccclccccc}
\hline
\multirow{2}{*}{Data} & \multicolumn{3}{c}{Pre-train} &  & \multicolumn{5}{c}{Zero-shot/Fine-tune}                 \\ \cline{2-4} \cline{6-10} 
                      & India    & Spain   & Canada   &  & Australia & Mexico & Germany & Japan    & Online Retail \\ \hline
Laguage               & English  & Spain   & English  &  & English   & Spain  & English & Japanese & English       \\
\#Items               & 45893    & 39675   & 99376    &  & 42094     & 43095  & 70527   & 22591    & 4223          \\
\#Users               & 507581   & 400883  & 992366   &  & 86975     & 249229 & 997555  & 277570   & 24446         \\
\#Inters              & 748607   & 708103  & 1971956  &  & 213086    & 483660 & 1840912 & 465746   & 540455        \\
\#Unseen              & /        & /       & /        &  & 14525     & 32477  & 60637   & 12831    & 516009        \\ \hline
\end{tabular}}
\caption{Dataset statistics for different domains of XMarket and Online Retail. Language indicates the default language used on Online Retail and the Amazon website for that domain (country). \#Inters is the number of interactions, \#Unseen is the number of target-domain samples that are non-overlapping with source domains.}
\vskip -0.5cm
\label{table:statistic}
\end{table*}

\section{Additional Training Details}
We use one Tesla V100 GPU for training the model. The pre-training phase takes around 25 minutes per epoch, and the maximum number of training epochs is 30.

For contrastive learning, \unisrec uses in-batch negative. Thus, the negative items are from all domains. \model adopts a more complex implementation, sampling negative items randomly and uniformly from the same domain as the positive item.

\begin{table*}
\resizebox{0.45\textwidth}{!}{
\begin{tabular}{ll}
\toprule
hyperparameter                                                                      & candidate value              \\ \midrule
learning rate                                                                       & 0.00003, \textbf{0.0003}, 0.003, 0.03 \\
batch size                                                                          & 32, 64, 128, \textbf{256}, 512        \\
\# of negtive samples                                                               & 63, 127, \textbf{255}, 511            \\
\# of gru layers                                                                    & 1, \textbf{2}, 3, 4, 6, 8, 12         \\
\# of attention layers                                                              & 1, \textbf{2}, 3, 4, 6, 8, 12         \\
\# of attention heads                                                               & 1, \textbf{2}, 3, 4, 6, 8, 12         \\
\begin{tabular}[c]{@{}l@{}}\# of embedding dimensions\\  for user/item\end{tabular} & 64, 128, \textbf{256}, 512, 1024      \\
domain bias regularization                                                          & 0.03, \textbf{0.3}, 3, 30, 300        \\
item bias regularization                                                            & \textbf{100}                         \\ \bottomrule
\end{tabular}}
\caption{The hyperparameter tuning setting. The selected ones are marked in \textbf{bold face}.}
\vskip -0.5cm
\label{table:hyp}
\end{table*}

For hyperparameter tuning, we consider candidates as shown in Table~\ref{table:hyp}.

\end{document}

%% file: defs.tex
\def\D{{\bf D}}

\def\F{{\bf F}}

\def\H{{\bf H}}

\def\I{{\bf I}}

\def\R{{\bf R}}
\def\X{{\bf X}}

\def\S{{\bf S}}

\def\x{{\bf x}}

\def\Z{{\bf Z}}

\def\m{{\bf m}}
\def\n{{\bf n}}
\def\U{{\bf U}}
\def\u{{\bf u}}
\def\V{{\bf V}}
\def\v{{\bf v}}
\def\W{{\bf W}}

\def\0{{\bf 0}}
\def\1{{\bf 1}}

\def\ep{\mbox{\boldmath$\epsilon$\unboldmath}}

\def\muu{\mbox{\boldmath$\mu$\unboldmath}}

\newcommand{\secref}[1]{Sec.~\ref{#1}}

\newcommand{\figref}[1]{Fig.~\ref{#1}}

\newcommand{\eqnref}[1]{Eqn.~\ref{#1}}

\renewcommand{\frac}{\tfrac}

\renewcommand{\hat}{\widehat}